\documentclass[camera]{jpaper}

\usepackage{amsmath,amssymb}
\usepackage{graphicx}
\usepackage{url}
\usepackage{fancyhdr}
\usepackage{xspace}
\usepackage{graphicx}
\usepackage{subcaption}
\usepackage{algorithm}
\usepackage{algorithmic}
\usepackage{wrapfig}
\usepackage{tikz}
\usepackage{amsmath}
\usepackage{titlesec}
\usepackage[nocompress]{cite}
\usepackage{setspace}
\usepackage[bottom]{footmisc}
\usepackage{siunitx}
\usepackage{indentfirst}
\usepackage{authblk}
\usepackage{fixltx2e}
\usepackage[nolessnomore, italic]{mathastext}
\usepackage[T1]{fontenc}
\usepackage[usenames,dvipsnames,svgnames,table]{xcolor}
\usepackage[normalem]{ulem}
\usepackage{enumitem}
\usepackage[us,12hr]{datetime}
\usepackage[keeplastbox]{flushend}
\usepackage[hidelinks]{hyperref}

\makeatletter
\let\MYcaption\@makecaption
\makeatother

\usepackage[font=footnotesize]{subcaption}

\makeatletter
\let\@makecaption\MYcaption
\makeatother

\newcommand*\circled[1]{\tikz[baseline=(char.base)]{
            \noindent
            \node[shape=circle,draw,fill=black,text=white,inner sep=0.5pt] (char) {#1};}}

\newcommand{\ignore}[1]{}

\newcommand{\note}[1]{}
\newcommand{\squishlist}{
   \begin{list}{$\bullet$}
    { \setlength{\itemsep}{0pt}    \setlength{\parsep}{0pt}
      \setlength{\topsep}{0pt}     \setlength{\partopsep}{0pt}
      \setlength{\leftmargin}{2em} \setlength{\labelwidth}{1.5em}
      \setlength{\labelsep}{0.5em} } }

\newcommand{\squishend}{
    \end{list} }

\newif\ifcameraready
\camerareadytrue

\newcommand{\versionnum}[0]{5.1}

\ifcameraready
\else
\fi

\ifcameraready
  \newcommand{\todo}[1][]{}
  \newcommand{\ch}[0]{}
  \newcommand{\chs}[0]{}
  \newcommand{\chI}[0]{}
\else
  \newcommand{\todo}[1][]{\textbf{\fcolorbox{black}{red}{\color{white}{TODO}}} \underline{$\overline{\hbox{\emph{#1}}}$}}
  \newcommand{\ch}[0]{}
  \newcommand{\chs}[1]{{\color{cyan} #1}\xspace}
  \newcommand{\chI}[1]{{\color{BrickRed} #1}}
\fi

\begin{document}

\title{Exploiting Row-Level Temporal Locality in DRAM \\ to Reduce the Memory Access Latency}

\author{
{Hasan Hassan$^{1,2,3}$}%
\qquad%
{Gennady Pekhimenko$^{4,2}$}%
\qquad%
{Nandita Vijaykumar$^2$}%
\qquad%
{Vivek Seshadri$^{5,2}$}%
\vspace{2pt}\\%
{Donghyuk Lee$^{6,2}$}%
\qquad%
{Oguz Ergin$^3$}%
\qquad%
{Onur Mutlu$^{1,2}$}%
}
\affil{
{\em
$^1$ETH Z{\"u}rich\qquad 
$^2$Carnegie Mellon University\qquad
$^3$TOBB University of Economics \& Technology}%
\vspace{2pt}\\%
{\it
$^4$University of Toronto\qquad 
$^5$Microsoft Research India\qquad 
$^6$NVIDIA Research}%
}

%

\maketitle


\begin{abstract}

\chs{This paper summarizes the idea of ChargeCache, which was published in HPCA
2016~\cite{hassan2016chargecache}, and examines the work's significance and
future potential.} 
DRAM latency continues to be a critical bottleneck for
system performance. In this work, we develop a low-cost mechanism, called
\emph{ChargeCache}, that enables faster access to recently-accessed rows in
DRAM, with no modifications to DRAM chips. Our mechanism is based on the key
observation that a recently-accessed row has more charge and thus the following
access to the same row can be performed faster. To exploit this observation, we
propose to track the addresses of recently-accessed rows in a table in the
memory controller. If a later DRAM request hits in that table, the memory
controller uses lower timing parameters, leading to reduced DRAM latency. Row
addresses are removed from the table after a specified duration to ensure rows
that have leaked too much charge are not accessed with lower latency. We
evaluate ChargeCache on a wide variety of workloads and show that it provides
significant performance and energy benefits for both single-core and multi-core
systems.

\end{abstract}


\section{Problem: DRAM Latency}

DRAM technology is commonly used as the main memory of modern computer
systems. This is because DRAM is at \ch{a} more favorable point in the
trade-off spectrum of density (cost-per-bit) and access latency
compared to other technologies like SRAM or flash. However, commodity
DRAM devices are heavily optimized to maximize cost-per-bit. In fact,
the latency of commodity DRAM has not reduced significantly in the
past \ch{two} decade\ch{s~\cite{lee,mutluresearch, chang2016understanding,
lee2016reducing, chang2017thesis, lee2015}}.

The latency of DRAM is heavily dependent on the design of the
DRAM chip architecture, specifically the length of a wire called
\emph{bitline}. A DRAM chip consists of millions of DRAM cells.
Each cell is composed of a transistor-capacitor pair. To access
data from a cell, DRAM uses a component called \emph{sense
amplifier}. Each cell is connected to a sense amplifier using a
\emph{bitline}. To amortize the large cost of the sense
amplifier, hundreds of DRAM cells are connected to the same
bitline~\cite{lee}. \ch{A longer bitline leads} to \ch{higher}
resistance and parasitic capacitance on the path between \ch{a} DRAM
cell and the sense amplifier. As a result, longer bitlines result
in higher DRAM access latency~\ch{\cite{lee, lee2015, son,
lee2016reducing}}.

To mitigate the negative effects of long DRAM access latency, existing
systems rely on several major approaches. First, they employ large
on-chip caches to exploit the temporal and spatial locality of memory
accesses. However, cache capacity is limited by chip area. Even caches
as large as tens of megabytes may not be effective for some
applications due to very large working sets and memory access
characteristics that are not amenable to
caching~\cite{palacharla1994evaluating,qureshi2007adaptive,jevdjic2014unison,
  lotfi2012scale, qureshi2007line}. Second, systems employ aggressive
prefetching techniques to preload data from memory before it is
needed~\cite{baer1991effective, srinath2007feedback,
  charney1997prefetching}. However, prefetching is inefficient for
many irregular access patterns and it increases the bandwidth
requirements and interference in the memory
system~\ch{\cite{ebrahimi2011prefetch, ebrahimi2009techniques,
  ebrahimi2009coordinated,lee2008prefetch, srinath2007feedback,
  seshadri2015mitigating}}. Third, systems employ
multithreading~\ch{\cite{thornton1964parallel, smith1978pipelined,
lindholm2008nvidia}}.
However, this approach increases contention in the memory
system~\cite{moscibroda2007memory, ebrahimi2011parallel, mutlu08,
  das2013application} and does not aid single-thread
performance~\cite{joao2012bottleneck,
  suleman2009accelerating}. Fourth, systems exploit memory level
parallelism~\cite{mutlu2003runahead, glew1998mlp,
  chou2004microarchitecture, mutlu2005techniques, mutlu08}. The DRAM
architecture provides various levels of parallelism that can be
exploited to simultaneously process multiple memory requests generated
by modern processor
architectures~\cite{tomasulo1967efficient,patt1985hps,mutlu2003runahead,lee2009improving}.
While prior works\ch{~\cite{jeong12, mutlu08, ding04, lee2009improving,
  pai1999code, chou2004microarchitecture}} \ch{propose} techniques to better utilize the available
parallelism, the benefits of these techniques are limited due to
1)~address dependencies \ch{between} instructions in the
programs~\cite{avd,clap,liupp}, and 2)~resource conflicts in the
memory subsystem~\cite{kim12,rau1991pseudo}. Unfortunately,
\emph{none} of these four approaches \emph{fundamentally} reduce
memory latency at its \emph{source} and the DRAM latency continues to be a
performance bottleneck in many systems.

\section{Existing Techniques That Reduce \\ DRAM Latency}

DRAM latency can be reduced using several techniques\ch{, all of} which
have their own \ch{specific} shortcomings. One simple approach to reduce
DRAM latency is to use shorter bitlines. In fact, some specialized DRAM
chips~\cite{rldram, lldram, sato1998fast} offer lower latency by using
shorter bitlines compared to commodity DRAM chips. Unfortunately, such
chips come at a significantly higher cost \ch{than chips that use long
bitlines,} as they reduce the overall density of the device because they
require more sense amplifiers, which occupy significant area~\cite{lee}.
Therefore, such specialized chips are usually not desirable for systems
that require high memory capacity~\cite{chatterjee2012leveraging}. Prior
works have proposed several heterogeneous DRAM architectures (e.g.,
segmented bitlines~\cite{lee}, asymmetric bank
organizations~\cite{son}\ch{, mechanisms that exploit the inherent latency
variation across cells~\cite{chang2016understanding, lee2017design}})
that divide DRAM into two regions: one with low latency, and another with
slightly higher latency.  Such schemes propose to map frequently accessed
data to the low-latency region, thereby achieving lower average memory
access latency. However, such schemes \ch{might} require 1)~non-negligible changes to
the cost-sensitive DRAM design, \ch{2)~techniques to create or identify
low-latency regions in DRAM, \chs{and/or} 3)}~mechanisms to identify, map, and
migrate frequently-accessed data to low-latency regions. As a result, even
though they reduce the latency for some portions of the DRAM chip, they may
\chI{not be easy} to adopt.



\section{Key Observations}

\chI{In our HPCA 2016 paper~\cite{hassan2016chargecache}, we make two major 
observations that motivate a new mechanism for reducing DRAM latency,}

\textbf{Charge Variation.} The amount of charge in \ch{the} DRAM cells
\ch{of a row} determines the required latency for a DRAM access \ch{to that
row}. If the amount of charge in the cell is low, the sense amplifier
completes its operation in longer time. Therefore, DRAM access latency
increases. A DRAM cell loses its charge over time and the charge is
replenished by a refresh operation or an access to the row.  The access
latency of a cell whose charge has been replenished recently can thus be
significantly lower than the access latency of a cell that has less charge.
Our SPICE simulations show that the first read/write command can be issued
44\% faster to a highly-charged DRAM row compared to a row with less
charge\ch{~(see Section~\ref{subsection:spice_sim} and our HPCA 2016
paper~\cite{hassan2016chargecache})}.

\textbf{Row-Level Temporal Locality.} We find that, mainly due to DRAM bank
conflicts~\cite{kim12, rau1991pseudo}, many applications tend to access
rows that were recently closed (i.e., closed within a very short time
interval). We refer to this form of temporal locality where certain rows
are \ch{frequently} closed and \ch{re-opened} as \textit{Row-Level Temporal
Locality (RLTL)}.  An important outcome of this observation is that a DRAM
row remains in a \emph{highly-charged} state when accessed for the
\emph{second} time within a short interval after the prior access. This is
because accessing the DRAM row inherently replenishes the charge within the
DRAM cells (just like a refresh operation does)~\cite{liu2012raidr, shin,
nair2013, chang2014, ghosh2007smart, liu2013experimental}.

\begin{figure}[ht]
	\centering
	\includegraphics[width=.95\linewidth]{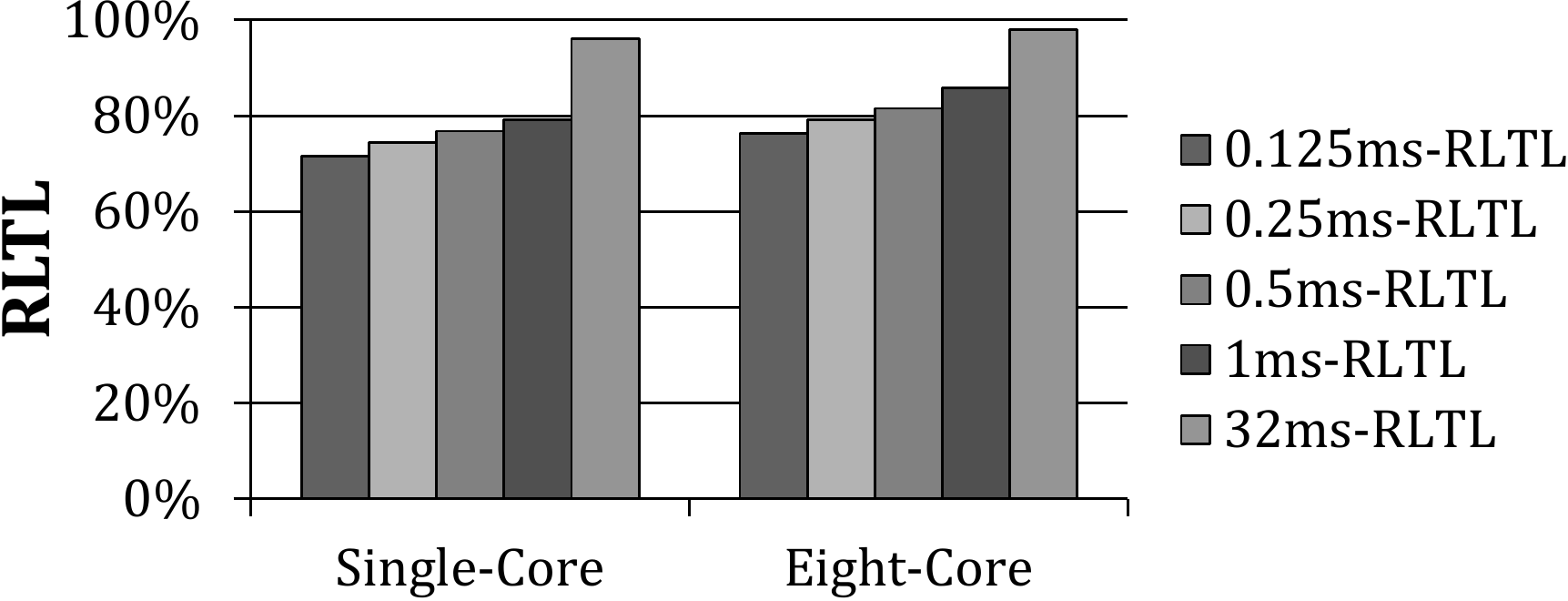}
    \caption{Average \ch{row-level temporal locality (RLTL)} for \ch{22}
    single-core and \ch{20} \chs{eight-core} workloads.}
	\label{figure:rltl}
\end{figure}

We define \emph{t-RLTL} of an application for a given time interval
\emph{t} as the fraction of row activations in which the activation occurs
within the time interval \emph{t} after a \emph{previous} precharge to the
same row. Figure~\ref{figure:rltl} shows \ch{the} average RLTL for
single-core and \chs{eight-core} workloads with five different time intervals (from
$0.125ms$ to  $32ms$). \ch{Our detailed experimental methodology is
described in Section 5 of \chs{our HPCA 2016 paper}~\cite{hassan2016chargecache}.} For single-core
workloads\ch{,} the average \ch{$1ms$-RLTL is 83\%}. In other words,
\ch{83\%} of
all the row activations occur within \ch{$1ms$} after the \ch{same} row was
previously precharged. Due to the additional bank conflicts \ch{incurred
as the number of workloads executing increases}, for \chs{eight-core} workloads\ch{,} the
average \ch{$1ms$-RLTL is 89\%}, significantly higher than that for the
single-core workloads. These results show that RLTL of both single-core and
\chs{eight-core} workloads is significantly high even for small \ch{values of}
\emph{t}, \chI{motivating} us to exploit RLTL\ch{~(i.e., row-level temporal
locality)} to detect highly-charged DRAM rows.\ch{\footnote{For a more
detailed study of row-level temporal locality, please see Section~3
of our HPCA 2016 paper~\cite{hassan2016chargecache}.}}

\ch{Note that a major reason for the high row-level temporal locality is
the occurrence of bank conflicts in the DRAM subsystem. We find that, due to
the bank conflicts, a row is likely to be requested again soon after it is
precharged due to an \chs{intervening} request to the same bank.}

\section{Our Goal}

We observe that \emph{many} applications exhibit high row-level
temporal locality. In other words, for many applications, a
significant fraction of the row activations occur within a small
interval after the corresponding rows are precharged. As a
result, such row activations can be served with lower activation
latency than specified by the DRAM standard. \textbf{Our goal} in
this work is to exploit this observation to reduce the effective
DRAM access latency by tracking recently-accessed DRAM rows in
the memory controller and reducing the latency for their next
access(es). To this end, we propose an efficient mechanism,
ChargeCache, which we describe in the next section.

\section{Solution: ChargeCache}

ChargeCache is based on three observations: 1)~\ch{a row whose cells'
charge has been}
recently \ch{replenished} can be accessed with lower activation latency,
2)~activating a row \ch{replenishes} the charge on the cells of that row
and the cells start leaking only after the \emph{following} precharge
command, and 3)~many applications exhibit high row-level temporal
locality, i.e., recently-activated rows are more likely to be
activated again. Based on these observations, ChargeCache tracks
rows that are recently activated, and serves \ch{near-future} activates to
such rows with lower latency by lowering the DRAM timing
parameters for such activations.

\begin{figure} [h!]
    \centering
    \includegraphics[width=.95\linewidth]{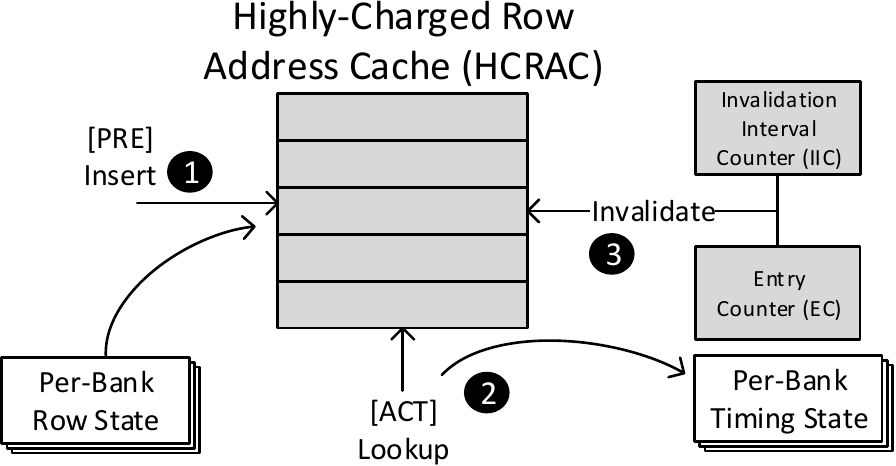}
    \caption{Components of the ChargeCache Mechanism. Reproduced
    from~\cite{hassan2016chargecache}\chs{.}}
    \label{figure:charge_cache}
\end{figure}

As we show in Figure~\ref{figure:charge_cache}, ChargeCache adds a small
table (\ch{structured as a} cache), called \emph{High-Charged Row Address Cache (HCRAC)}, to
the memory controller that tracks the addresses of recently-accessed DRAM
rows, i.e., highly-charged rows. ChargeCache performs three operations.
First, when a precharge command is issued to a bank, ChargeCache inserts
the address of the row that was activated in the corresponding bank to the
table \chs{(\circled{1} in the figure)}. Second, when an activate command is issued, ChargeCache checks if
the corresponding row address is present in the table \chs{(\circled{2})}. If the address is
\emph{not} present, then ChargeCache uses the standard DRAM timing parameters to
issue subsequent commands to the bank. However, if the address of the
activated row \emph{is} present in the table, ChargeCache employs reduced timing
parameters for subsequent commands to that bank. Our experimental results
on multi-programmed applications show that, on average, ChargeCache can
reduce the latency of 67\% of all DRAM row activations\ch{~(as shown in
Section 6.4 of \chs{our HPCA 2016 paper}~\cite{hassan2016chargecache}).} Third, ChargeCache
periodically invalidates old entries from the table to ensure that only rows
that have sufficient amount of charge for being accessed with low latency
remain in the table \chs{(\circled{3})}. Since a row may potentially reside in the
table for very long time without being activated, such an operation is necessary
to avoid a low-latency access to a row with small amount of
charge\ch{~(which could lead to wrong results)}.

We \chs{name} \ch{our} mechanism \emph{ChargeCache}\chs{,} as it provides a
\emph{cache}-like benefit, i.e., latency reduction based on a
locality property (i.e., RLTL), and does so by taking advantage
of the \emph{charge} level stored in a recently-activated row.
The mechanism could potentially be used with current and emerging
DRAM-based memories where the stored charge level leads to
different access latencies. We release the source code of
ChargeCache for two different versions of
Ramulator~\cite{ramulator, ramulator_web, ramulatorSharp_web} to
enable future research to build upon our ideas.

\section{\ch{Experimental} Evaluation}
\label{sec:evaluation}

In this section, we first explain our experimental methodology. Later, we
quantitatively analyze the \ch{system} performance improvement and
\ch{DRAM} energy savings that ChargeCache provides. 

\subsection{Methodology}
\label{subsec:methodology}

We use circuit-level SPICE simulations to evaluate the DRAM latency
reduction that can be achieved when accessing a highly-charged DRAM row.
In Section~\ref{subsection:spice_sim}, we show the reduction in two DRAM timing
parameters, \textit{tRCD} and \textit{tRAS}, that are affected by high
charge amount stored in a DRAM cell.\ch{\footnote{For detail on DRAM timing
parameters and operation, we refer the reader to \chs{our prior
works}~\cite{kim12, lee, lee2015, hassan2016chargecache, seshadri2013,
chang2016understanding, chang2017understanding, kim2010thread, kim2010,
liu2012raidr, lee2017design, seshadri2017ambit, liu2013experimental, chang2016,
hassan-hpca2017, ramulator, patel2017reach, kim-hpca2018}.}}

To evaluate the performance of ChargeCache, we use a cycle-accurate DRAM
simulator, Ramulator~\cite{ramulator, ramulator_web}, in CPU-trace-driven
mode. CPU traces are collected using a Pintool~\cite{pintool}.
Table~\ref{table:system_config} lists the configuration of the evaluated
systems. We implement the \ch{HCRAC} table, which ChargeCache uses to store the
addresses of recently accessed DRAM rows, similarly to a 2-way associative
cache that uses the LRU policy.

\begin{scriptsize} 
    \begin{table}[h!] 
        \caption{Simulated system configuration\chs{.}  Reproduced
    from~\cite{hassan2016chargecache}.} 
        \centering 
        \renewcommand{\arraystretch}{1.4}
        \begin{tabular}{m{2cm} m{5cm}} 
            \hline 
            Processor & 1-8 cores, 4GHz clock frequency, 3-wide issue, 8
            MSHRs/core, 128-entry instruction window\\ 
            \hline 
            Last-level Cache & 64B cache-line, 16-way associative, 4MB
            cache size \\ 
            \hline 
            Memory \hspace{0.5mm} Controller & 64-entry read/write request
            queues, FR-FCFS scheduling policy~\cite{frfcfs,
            zuravleff1997controller}, open/closed row
            policy~\cite{kim2010thread, kim2010} for single/multi core\\ 
            \hline
            DRAM & DDR3-1600~\cite{micronDDR3}, 800MHz bus frequency, 1/2
            channels, 1 rank/channel, 8 banks/rank, 64K rows/bank, 8KB
            row-buffer size, tRCD/tRAS  11/28 cycles\\ 
            \hline 
            ChargeCache & 128-entry (672 bytes)/core, 2-way associativity,
            LRU replacement policy, $1ms$ \textit{caching duration},
            tRCD/tRAS reduction 4/8 cycles \\ 
            \hline 
        \end{tabular} 
        \label{table:system_config} 
    \end{table}
\end{scriptsize}

For area, power, and energy measurements, we modify
McPAT~\cite{mcpat} to implement ChargeCache using
$22 nm$ process technology. We use
DRAMPower~\cite{drampower} to obtain power/energy results \chs{for} the
off-chip main memory subsystem. We feed DRAMPower with DRAM
command traces obtained from our simulations using Ramulator.

We run 22 workloads from \chs{the} SPEC CPU2006~\cite{spec2006}, TPC~\cite{tpc}\chs{,}
and STREAM~\cite{stream} benchmark suites. We use
SimPoint~\cite{simpoint} to obtain traces from representative
phases of each application. For single-core evaluations, unless
stated otherwise, we run each workload for 1 billion
instructions. For multi-core evaluations, we use 20 \chs{multiprogrammed}
workloads by assigning a randomly-chosen application to each
core. We evaluate each configuration with its \ch{best-performing}
row-buffer management policy. Specifically, we use the open-row
policy for single-core and closed-row policy for multi-core
configurations. We simulate the benchmarks until each core
executes at least 1 billion instructions. For both \ch{single-} and
multi-core configurations, we first warm up the caches
and ChargeCache by fast-forwarding 200 million cycles.

We measure performance improvement for single-core workloads using
the Instructions per Cycle (IPC) metric. We measure multi-core
performance using the weighted speedup~\cite{snavely2000symbiotic}
metric. Prior work has shown that weighted speedup is a measure of
\ch{system-level job} throughput~\cite{eyerman2008system}.

\subsection{Reduction in DRAM Timing Parameters}
\label{subsection:spice_sim}

\begin{figure} [h!] 
    \centering 
    \includegraphics[width=.95\linewidth]{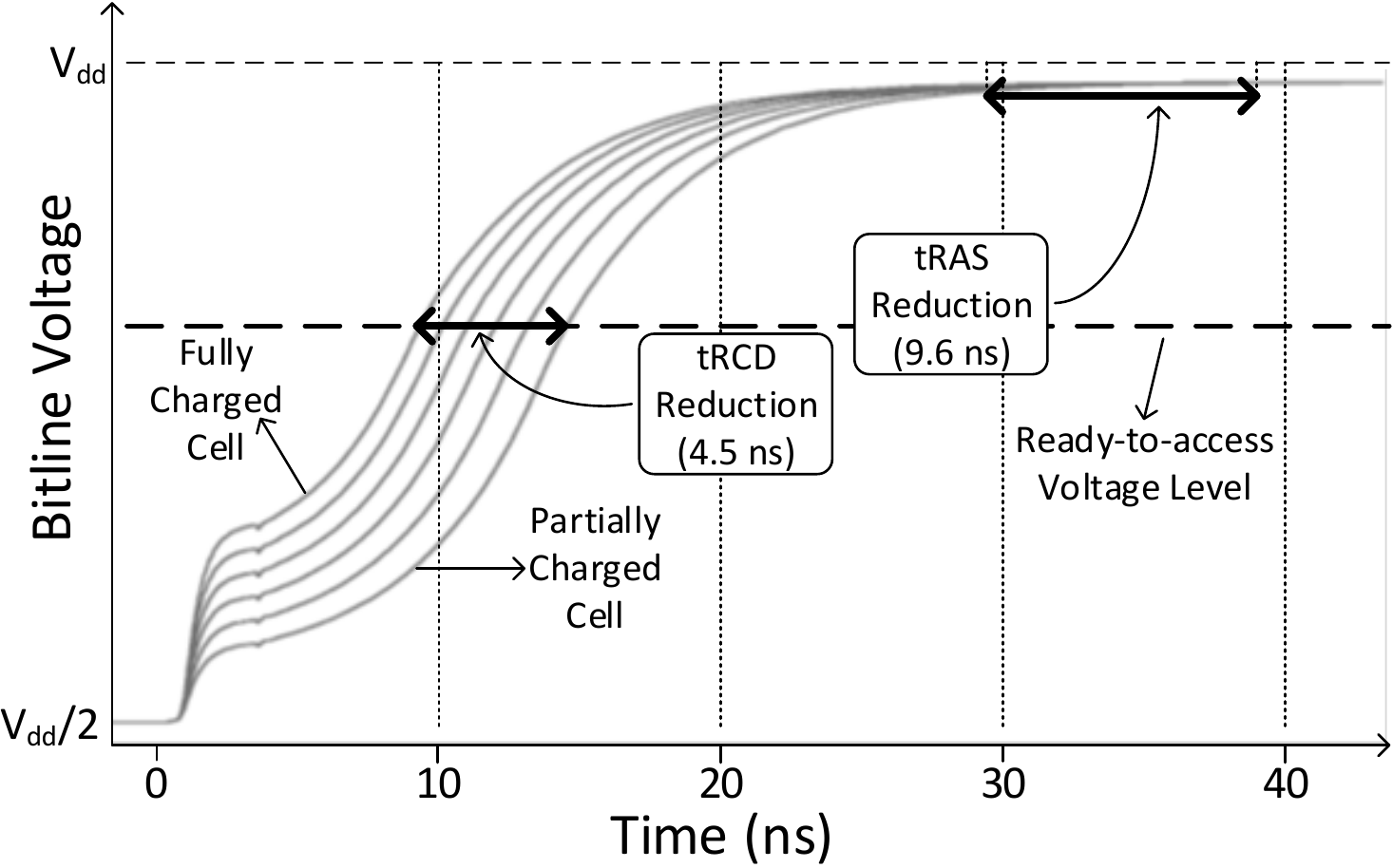}
    \caption{Effect of initial cell charge on bitline voltage. Reproduced
    from~\cite{hassan2016chargecache}\ch{.}}
    \label{figure:spice} 
\end{figure}

We evaluate the potential reduction in \textit{tRCD} and \textit{tRAS} for
ChargeCache using circuit-level SPICE simulations. We implement the DRAM sense
amplifier circuit using $55nm$ DDR3 model
parameters~\cite{rambus} and PTM low-power transistor models~\cite{zhaoptm,
ptmweb}. Figure~\ref{figure:spice} plots the variation in bitline voltage level
during cell activation for different initial charge amounts of the cell.


Depending on the initial charge (i.e., voltage level) of the cell, the
bitline voltage increases at different speeds. When the cell is
\emph{fully-charged}, the sense amplifier is able to drive the bitline
voltage to the \emph{ready-to-access voltage level} in only
$10ns$. However, a partially-charged cell (i.e., one
that has not been accessed for $64ms$) brings the bitline voltage up
slower. Specifically, the bitline connected to such a
partially-charged cell reaches the ready-to-access voltage level in
$14.5ns$. Since DRAM timing parameters are dictated
by this worst-case partially-charged state right before the refresh
interval, we can achieve \chs{a} $4.5ns$ reduction in
\textit{tRCD} for a \emph{fully-charged} cell.  Similarly, the charge of the
cell capacitor is restored at different times depending on the initial
voltage of the cell. For a fully-charged cell, this results in
\chs{a} $9.6ns$ reduction in \textit{tRAS}.


In practice, we expect DRAM manufacturers to identify the
lowered timing constraints for different caching durations.
Today, DRAM manufacturers test each DRAM chip to determine if it
meets the timing specifications. Similarly, we expect the
manufacturers would also test each chip to determine if it meets
the ChargeCache timing constraints.

\subsection{Results}
\label{subsec:results}

We experimentally evaluate the following mechanisms: 1)
ChargeCache, 2) NUAT~\cite{shin}, which accesses \emph{only} rows
that are \emph{recently-refreshed} at lower latency than the DRAM
standard, 3) ChargeCache + NUAT, which is a combination of
ChargeCache and NUAT~\cite{shin} mechanisms, and 4) Low-Latency
DRAM (LL-DRAM)~\cite{rldram}, which is an idealized comparison
point where we assume \emph{all rows} in DRAM can be accessed
with low latency, compared to our baseline DDR3-1600 memory, at
any time, \ch{irrespective} of when they are accessed or refreshed.

We compare the performance of our mechanism against the most
closely related previous work, NUAT~\cite{shin}. The key
idea of NUAT is to access \emph{recently-refreshed} rows at low
latency, because these rows are already highly-charged. Thus,
NUAT does \ch{\emph{not} use low latency for} rows that are
recently-\emph{accessed}, and hence it does \ch{\emph{not}}
exploit \ch{the} RLTL (Row-Level Temporal Locality) present in
many applications.



\begin{figure*}[!ht]
\centering
\begin{subfigure}[b]{\linewidth}
    \centering
    \includegraphics[width=.98\linewidth]{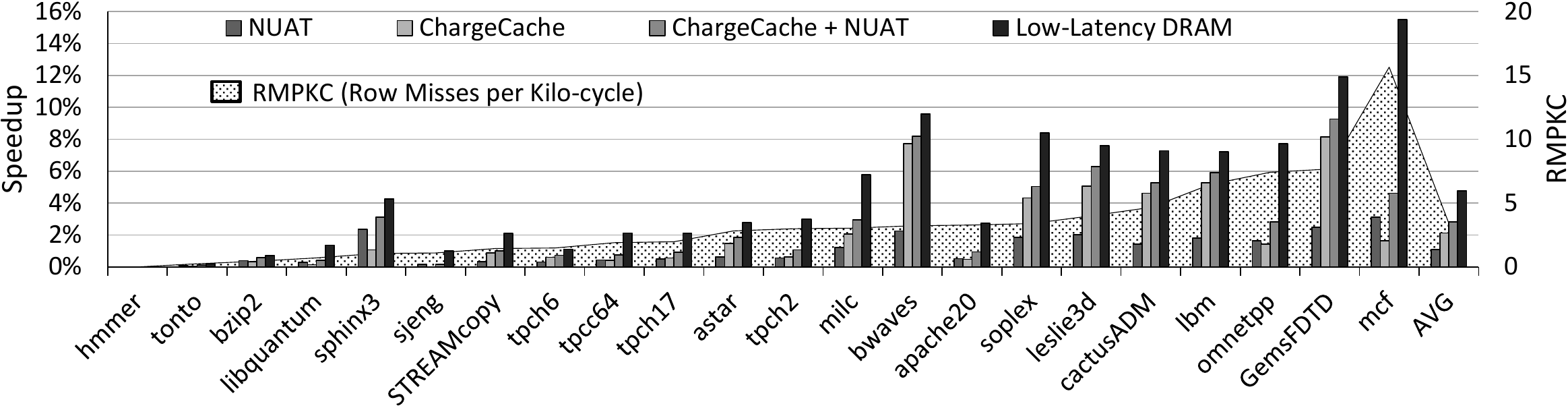}
    \vspace{-2mm}
    \caption{Single-core workloads}
    \vspace{1mm}
    \label{subfigure:ipc_sc}
\end{subfigure}
\begin{subfigure}[b]{\linewidth}
    \centering
    \includegraphics[width=.98\linewidth]{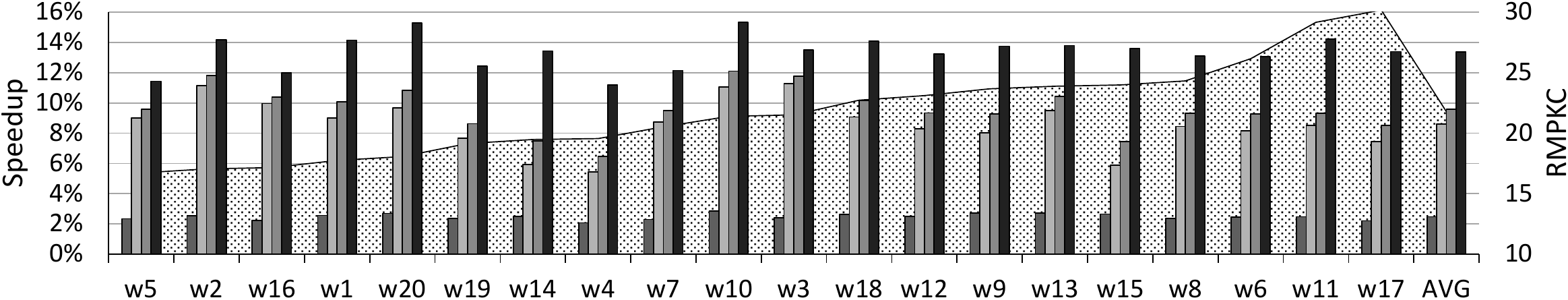}
    \vspace{-1mm}
    \caption{Eight-core workloads}
    \label{subfigure:ipc_8c}
\end{subfigure}
\caption{Speedup with
ChargeCache, NUAT and Low-Latency DRAM for single-core and
    eight-core workloads. Reproduced from~\cite{hassan2016chargecache}.}
\label{figure:ipc}
\end{figure*}

Figure~\ref{figure:ipc} shows the performance of single-core and
eight-core workloads. The figure also includes the number of row
misses per kilo-cycles (RMPKC) to show row activation intensity,
which provides insight into the RLTL of the workload.

\emph{Single-Core Performance:} Figure~\ref{subfigure:ipc_sc}
shows the performance improvement over the baseline system for
single-core workloads. These workloads are sorted in ascending
order of RMPKC. ChargeCache achieves up to 9.3\% (an
average of 2.1\%) speedup.
Our mechanism outperforms NUAT and achieves a speedup close to
LL-DRAM with a few exceptions. Applications that have a wide gap in
performance between ChargeCache and LL-DRAM (\chs{e.g.,} \emph{mcf,
omnetpp}) access a large number of DRAM rows and exhibit high
row-reuse distance~\cite{kandemir2015}. A high row-reuse distance
indicates that there is \ch{a} large number of accesses to other rows
between two accesses to the same row. Due to this reason, ChargeCache
\emph{cannot} retain the addresses of highly-charged rows until the next
access to that row. Increasing the number of ChargeCache entries
or employing cache management policies aware of reuse
distance or thrashing~\cite{duong2012, seshadri2012evicted,
qureshi2007adaptive, tyson1995modified} may improve the performance of
ChargeCache for such applications. We leave the evaluation of these
methods for future work. We conclude
that ChargeCache significantly reduces execution time for most
high-RMPKC workloads and outperforms NUAT for all but few
workloads.

\emph{Eight-Core Performance:} Figure~\ref{subfigure:ipc_8c} shows the
speedup on eight-core multiprogrammed workloads. On average,
ChargeCache and NUAT improve performance by 8.6\% and 2.5\%,
respectively. Employing ChargeCache in combination with NUAT
achieves a 9.6\% speedup, which is only 3.8\% less than the
improvement obtained using LL-DRAM. Although the multiprogrammed
workloads are composed of the \emph{same} applications as in
single-core evaluations, we observe much higher performance
improvements \ch{for the} eight-core workloads. The reason is twofold.
First, since multiple cores share a limited capacity LLC,
simultaneously running applications compete for the LLC. Thus,
individual applications access main memory more often, which
leads to higher RMPKC. This makes the workload
performance more sensitive to main memory
latency~\cite{chandra2005predicting, iyer2007qos, kim12}. Second, the
memory controllers receive memory requests from multiple
simultaneously-running applications to a limited number of memory
banks. Such requests are likely to
target different rows since they use separate memory regions and
these regions map to separate rows. Therefore, applications
running concurrently
exacerbate the bank-conflict rate and increase the number of row
activations that hit in ChargeCache.

Overall, ChargeCache improves performance by up to \ch{11.3\% (8.1\%)
and \chI{on average} 8.6\% (2.1\%) for eight-core (single-core)}
workloads. It outperforms NUAT for most of the applications\ch{. Using}
NUAT in combination with ChargeCache improves ch{system}
performance \ch{even} further.

\subsection{Impact on DRAM Energy}
\label{subsection:energy_efficiency}

\begin{figure}[!ht] \centering
\includegraphics[width=.85\linewidth]{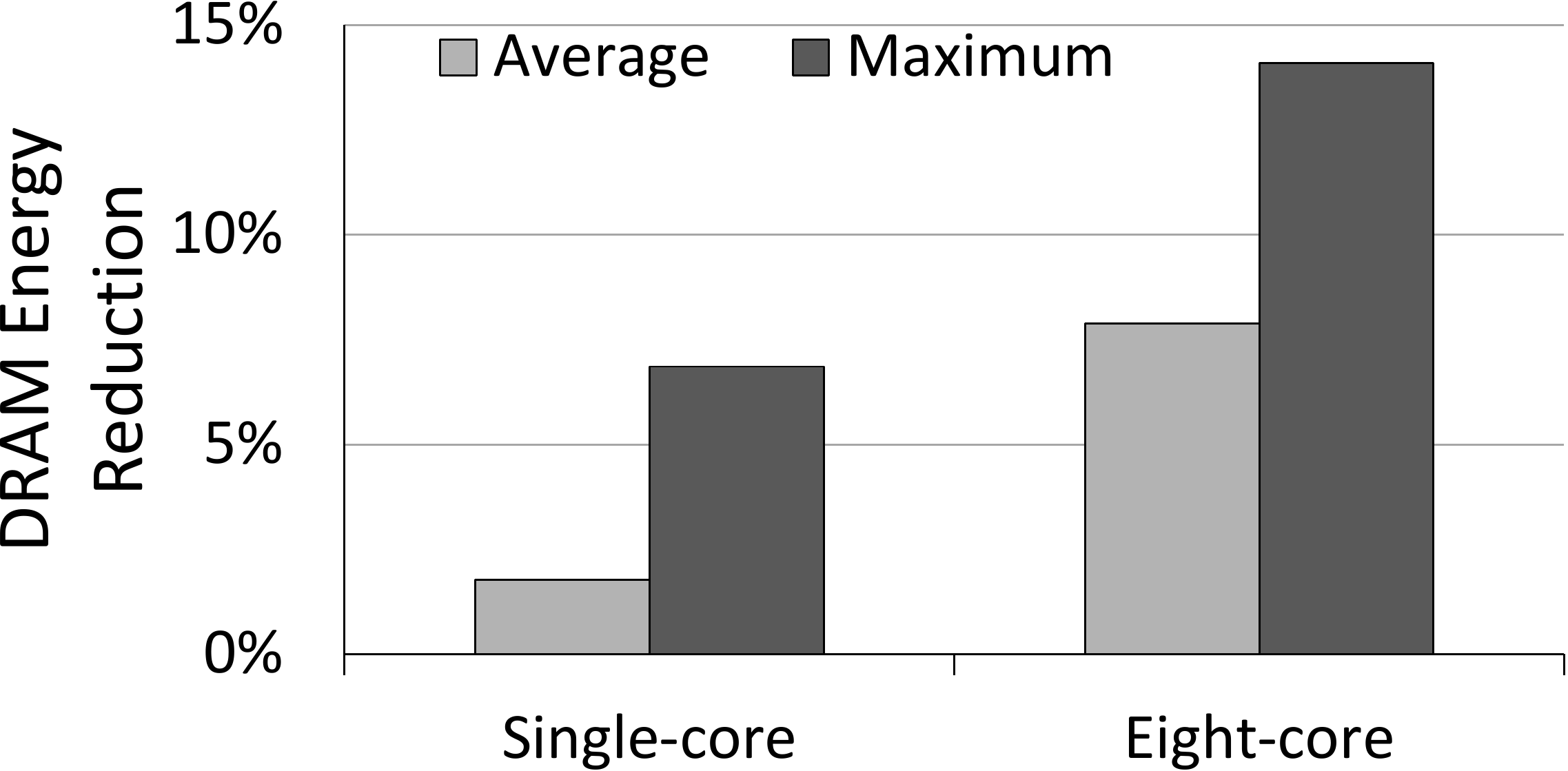}
    \caption{DRAM energy reduction of ChargeCache. Reproduced
    from~\cite{hassan2016chargecache}\chs{.}}
\label{figure:energy}
\end{figure}

ChargeCache incurs negligible area and power overheads
(\ch{see} Section~\ref{subsection:area_power_overhead}). Because it
reduces execution time with negligible overhead, it leads to
significant energy savings. Even though ChargeCache increases the
energy efficiency of the entire system, we quantitatively evaluate
the energy savings only for the DRAM subsystem since
Ramulator~\cite{ramulator} \ch{currently} does not have a detailed CPU model.
Figure~\ref{figure:energy} shows the average and maximum DRAM
energy savings for single-core and eight-core workloads.
ChargeCache reduces energy consumption by \chs{an average of 7.9\% (1.8\%), and by}
up to \ch{14.1\% (6.9\%)\chs{,} for eight-core (single-core)}
workloads. We conclude that ChargeCache is effective at improving
the energy efficiency of the DRAM subsystem, as well as the
entire system.

\subsection{Area and Power Consumption Overhead}
\label{subsection:area_power_overhead}

HCRAC (Highly-Charged Row Address Cache) is the most area/power
demanding component of ChargeCache. 
As we replicate \ch{HCRAC} on a per-core and per-memory
channel basis, the total area and power overhead ChargeCache
introduces depends on the number of cores and memory
channels.\footnote{Note that sharing \ch{a single HCRAC} across \ch{all or
multiple} cores can
  result in even lower \ch{overhead}. We leave the exploration of such
  \ch{shared-HCRAC}
designs to future work.} The total storage requirement is
given by
Equation~\ref{eq:2}, where \textit{C} are \textit{MC} are the
number of cores and memory channels, respectively.
\textit{LRUbits} depends on \ch{HCRAC} associativity. \textit{EntrySize} is
calculated using Equation~\ref{eq:1}, where \textit{R},
\textit{B}, and \textit{Ro} are the number of ranks, banks, and
rows in DRAM, respectively.

\vspace{-4mm}
\begin{equation}
\label{eq:2}
\resizebox{.91\linewidth}{!}{
$Storage_{bits} = C * MC * Entries * (EntrySize_{bits} + LRU_{bits})$}
\end{equation}

\vspace{-9mm}
\begin{equation} \label{eq:1}
 \resizebox{.82\linewidth}{!}{$Entry Size_{bits} = log_{2}(R) +
log_{2}(B) +
 log_{2}(Ro) + 1$} \end{equation}


\textbf{Area.} Our eight-core configuration has two memory
channels.
This introduces a total of 5376 bytes in storage requirement
for a 128-entry \ch{HCRAC}, corresponding to an area of
\SI{0.022}{\milli\meter\squared}. This overhead
is only 0.24\% of the 4MB LLC.

\textbf{Power Consumption.} \ch{HCRAC} is accessed on every
\textit{activate} and \textit{precharge} command issued by the memory
controller. On an \emph{activate} command, \ch{HCRAC} is searched for
the corresponding row address. On a \emph{precharge} command, the
address of the precharged row is inserted into
\ch{HCRAC}. \ch{HCRAC} entries are periodically invalidated to
ensure they do not exceed a specified \textit{caching duration}. These
three operations increase dynamic power consumption in the memory
controller, and the \ch{HCRAC} storage increases static power
consumption.  Our analysis indicates that ChargeCache consumes
\SI{0.149}{\milli\watt} on average. This is only 0.23\% of the
average power consumption of the entire 4MB LLC. Note that we include
the effect of this additional power consumption in our DRAM energy
evaluations in Section~\ref{subsection:energy_efficiency}. We conclude
that ChargeCache incurs almost negligible chip area and power
consumption overheads.

\ch{
    \subsection{Other Results}

    We also evaluate and assess the sensitivity of ChargeCache benefits to
    ChargeCache capacity, caching duration, and temperature in Sections 6.4
    and 7.1 of \chs{our HPCA 2016 paper}~\cite{hassan2016chargecache}.

}


\section{Related Work}
\label{section:related_work}


To our knowledge, this paper is the first to (\textit{i}) show
that applications typically exhibit significant \textit{Row-level
Temporal Locality (RLTL)} and (\textit{ii}) exploit this locality
to improve system performance by reducing the latency of
requests to recently-accessed \ch{memory} rows.


We have already (in Section~\ref{subsec:results}) qualitatively and
quantitatively compared ChargeCache to NUAT~\cite{shin}, which reduces
access latency to \emph{only} recently-refreshed rows. We have \ch{also} shown that
ChargeCache \ch{provides} significantly higher average latency reduction than
NUAT because RLTL is usually high, whereas the fraction of accesses to rows
that are recently-refreshed is typically low~\ch{(\chs{see} Section~3
in \chs{our HPCA 2016 paper}~\cite{hassan2016chargecache})}.

Other previous works \ch{propose} techniques to reduce
performance degradation caused by long DRAM latencies. They
\ch{focus} on 1) enhancing the DRAM, 2) exploiting variations in
manufacturing process and operating conditions, 3) developing
\ch{various} memory scheduling policies. We briefly summarize how
ChargeCache differs from these works.

\textbf{Enhancing DRAM Architecture.} Lee at al.  propose Tiered-Latency
DRAM (TL-DRAM)~\cite{lee}\ch{,} which divides each subarray into near and far
segments using isolation transistors. With TL-DRAM, the memory controller
accesses the near segment with lower latency since the isolation transistor
reduces \ch{the} bitline capacitance in that segment. Our mechanism could be
implemented on top of TL-DRAM to reduce the access latency for both the
near and far segment. Kim et al. \ch{propose SALP, which unlocks}
parallelism \ch{between} subarrays at low
cost\ch{, \chI{by modifying the DRAM chip to} enable pipelined access
to subarrays}~\cite{kim12}. The goal of SALP is to
reduce the impact of bank conflicts \ch{by providing more parallelism and
thereby reducing the latency of bank-conflict accesses}. O et
al\chs{.}~\cite{son2014} propose a DRAM architecture where sense amplifiers are
decoupled from bitlines to mitigate precharge latency. Choi et
al\chs{.}~\cite{choi} propose to utilize multiple DRAM cells to store a single bit
when sufficient DRAM capacity is available. By using multiple cells, they
reduce activation, precharge and refresh latencies. Other
works\chI{~\cite{gulur2012, son, zhang2014, seshadri2013, seshadri2015fast,
seshadri2017ambit, seshadri2015gather, chang2014, chang2016, lee.pact15}} also propose
new DRAM architectures to lower DRAM latency \ch{for various types of
operations and accesses}.

\ch{
    \chs{Processing-in-memory (PIM)} architectures~\cite{ahn2015scalable, ahn2015pim,
hsieh2016accelerating, hsieh2016transparent, boroumand2017lazypim,
seshadri2017ambit, seshadri2013, seshadri2015gather, fraguela2003programming,
elliott1999computational, draper2002architecture, kang2012flexram,
kogge1994execube, oskin1998active, patterson1997case, shaw1981non,
stone1970logic, kim.bmc18, boroumand.asplos18} using 3D-stacked memory~\chs{\cite{lee2016simultaneous,
hbm2013, hmc2spec, loh}} reduce the \chI{\emph{observed latency, from the
perspective of the processor,} by moving \chs{some
computation} operations closer to DRAM. \chs{3D-stacked memories are well
suited for processing-in-memory due to their \chs{inclusion of a} logic layer, which allows \chs{for the}
efficient implementation of CMOS logic \chs{in DRAM} and offers high bandwidth to the DRAM
layers.} However, \chs{PIM architectures} do not fundamentally reduce the access latency of the
DRAM device, which ChargeCache does (for certain access patterns).
}}

Unlike ChargeCache, \chs{a large number of} these works require changes to the DRAM
architecture itself. The approaches taken by these works are largely
orthogonal \ch{to the ChargeCache approach} and ChargeCache could be
implemented together with any of these mechanisms to further \ch{reduce} the
DRAM latency.

\textbf{Exploiting Process and Operating Condition Variations.}
Recent studies~\cite{lee2017design, chang2016understanding,
chang2017understanding, lee2015, chandrasekar2014} \ch{propose} methods to
reduce the safety margins of the DRAM timing parameters when operating
conditions are appropriate (i.e., not worst-case). Unlike these works,
ChargeCache is largely independent of operating conditions like
temperature, as discussed in Section~\ref{subsec:long_term}, and is
orthogonal to these latency reduction mechanisms. 


\textbf{Memory Request Scheduling Policies.} Memory request scheduling
policies~(e.g.,\chs{~\cite{lee2010dram, frfcfs, mutlu2007stall, mutlu08,
kim2010, kim2010thread, subramanian2014blacklisting,
subramanian2015application, subramanian2013mise, usui2016dash,
subramanian2015blacklisting, ausavarungnirun2012staged,
moscibroda2008distributed, zuravleff1997controller, ipek2008self, mukundan2012morse, kaseridis2011minimalist, ghose2013improving, hur2004adaptive}}) reduce the average
DRAM access latency by improving DRAM parallelism, row buffer locality\chs{,} and
fairness in especially multi-core \ch{and heterogeneous} systems. 
ChargeCache can be employed in conjunction with the scheduling
policy that best suits the application and the underlying
architecture.

\section{Significance}

Main memory latency has \ch{a} critical impact on system
performance\ch{~\cite{mutlu2013memory}}.
Our \ch{work} proposes a new
low-cost mechanism to reduce DRAM latency\ch{, \emph{without}} any
modifications to the existing DRAM chip architecture. In this
section, we discuss the significance of our work by \ch{describing}
its novelty and \ch{expected} long-term impact. 

\subsection{Novelty}

ChargeCache reduces average DRAM latency by exploiting a type of DRAM
access locality, \emph{Row-Level Temporal Locality (RLTL)}, that
\ch{commonly exists in} workloads \ch{due to the presence of DRAM bank
conflicts}. \ch{Our work} is the first to \ch{observe and formally} define
\emph{RLTL} and exploit it to reduce DRAM latency by designing a \ch{new}
mechanism that \ch{takes advantage of} \emph{RLTL} and the fact that a DRAM
row gets inherently refreshed on access. Our mechanism does not require any
changes to the existing DRAM array structure \ch{of the DRAM chips} and can
be easily implemented on top of any DRAM standard with negligible overhead
in the memory controller logic.

\subsection{Applicability to Emerging DRAM Standards}

ChargeCache is applicable to any memory technology where cells are volatile
(leak charge over time) and the charge variation due to \ch{charge} leakage has
impact on access latency.  ChargeCache can be used with to a large set of
standards derived from DDR (DDRx, GDDRx, LPDDRx, etc.)~\cite{ramulator} in a
manner similar to the mechanism described in this work, without modifying the
DRAM architecture. Using ChargeCache with 3D-stacked
memories\ch{~\cite{lee2016simultaneous, loh}} such as \chs{Wide I/O}, HBM, and
HMC~\cite{ramulator} is also straightforward. The difference is that\chs{, for
the technologies that implement the memory controller in the logic layer,} the
DRAM controller, and hence ChargeCache, \ch{can} be \ch{easily} implemented in
the logic layer of the 3D-stacked memory chip instead of the processor chip.


\chI{We also believe that the key idea of ChargeCache is not limited to DRAM, 
and can potentially be applied to other memory technologies that store 
information in form of electrical charge, such as NAND flash 
memory~\cite{cai2017error, cai2015data, cai.bookchapter.arxiv17, cai.procieee.arxiv17,
cai2012flash, cai.date13, cai.date12, cai.dsn15, cai.sigmetrics14, cai.hpca17,
luo.jsac16, luo.hpca18}.}

\subsection{Long-Term Impact}
\label{subsec:long_term}

\subsubsection{Reducing DRAM Latency}
During the last \ch{several decades}, DRAM capacity increased significantly by
shrinking the feature size of the transistors. Similarly, more
efficient DRAM standards enabled memories with high bandwidth.
The new \ch{3D-stacking} technology offers even higher bandwidth by
incorporating DRAM and the \ch{logic layer on} the same chip \ch{in a
3D-stacked manner}. However,
none of these advances \ch{lead} to \ch{large improvements in the row
access latency of the DRAM arrays.} Hence, DRAM latency \ch{is already} a critical bottleneck for
system performance. Our work alleviates the DRAM latency problem 
with no overhead to the area-optimized DRAM chip\ch{,} which is
difficult to change\ch{, and with low overhead to the memory controller.}

\subsubsection{Row-Level Temporal Locality}
Our paper is the first work to observe \chs{row-level temporal locality (\emph{RLTL})}. Note that
\emph{RLTL} is different from Row-Reuse
Distance~\cite{kandemir2015} that a prior work studies. Row-Reuse
Distance is a metric \ch{indicating} the number of accesses
between two consecutive accesses to the same row.  On the other
hand, \emph{RLTL} indicates the \emph{time} between two
consecutive accesses to the same row. A \ch{row} locality metric that
includes time is important since \ch{charge leakage in DRAM is a function
of \emph{time}}. In this work, we exploit
\emph{RLTL} to reduce DRAM latency. However, \emph{RLTL} \ch{can} also
\ch{potentially} be used to discover new techniques to improve
\ch{different aspects of} DRAM\ch{,} such as
reliability\chI{~\cite{kim2014flipping, meza2015revisiting, mutlu2013memory, 
mutlu2017rowhammer}} and bandwidth.

\subsubsection{Importance for Future Systems} 
\ch{We believe the latency reduction mechanism of ChargeCache will become more
important in future systems for four reasons. First, DRAM latency will become a
much bigger bottleneck\chs{,} as applications will become more
\chs{data-intensive}~\cite{mutlu2013memory, mutluresearch}. Higher demand for
data will \chI{result} in more bank conflicts\chs{, as the number of DRAM banks is
not scaling as fast as data intensity}. Such applications will also have
\chI{fast} data access requirements, which will increase their sensitivity
to the memory access latency\chI{~\cite{mutlu2013memory, mutluresearch,
wang2014bigdatabench, ferdman2012clearing, kanev-isca2015}}. As bank conflicts increase and
accesses become more latency-critical, the benefits of ChargeCache will
increase\chs{,} as there will be \chI{higher} RLTL\chs{, which ChargeCache can exploit
to provide} higher performance improvement.

Second, ChargeCache is likely to remain much more
competitive than other state-of-the-art latency reduction techniques for
the 3D-stacked memories of the future.  \chI{These memories} will likely operate at higher
\ch{temperatures compared to conventional} DRAM chips.} 
\ch{The charge} leakage rate of DRAM
cells approximately doubles for every 10$^{\circ}$C increase in
temperature\ch{~\cite{lee2015, liu2013experimental, khan2014efficacy,
patel2017reach}}. This observation can be exploited to lower the DRAM latency
when operating at low temperatures. A previous study, \chs{Adaptive-Latency} DRAM
(AL-DRAM)~\cite{lee2015}, proposes a mechanism to improve system performance by
reducing the DRAM timing parameters at low operating temperature. \chs{AL-DRAM} is based
on the premise that DRAM typically does not operate at temperatures close to
the worst-case temperature (85$^{\circ}$~C) even when it is heavily accessed.
However, new 3D-stacked DRAM technologies such as HMC, HBM, WideIO may operate
at significantly higher temperatures due to tight integration of multiple stack
layers~\cite{black}. Therefore, \ch{state-of-the-art and compelling}
\textit{dynamic latency scaling} techniques such as AL-DRAM may be less useful
in these scenarios.
\ch{In contrast to AL-DRAM,} ChargeCache is \emph{not} based on the charge
difference that occurs due to temperature dependence. Rather, we exploit the
high level of charge in recently-precharged rows to reduce timing parameters
during later accesses to such rows. After conducting tests to determine the
possible latency reduction in accessing highly-charged rows (for ChargeCache
hits) at \emph{worst-case} temperatures, \ch{we show that} ChargeCache can be
employed independently of the operating temperature\ch{~(\chs{see} Section {7.1}
in \chs{our HPCA~2016 paper}~\cite{hassan2016chargecache})}. 

\ch{
Third, ChargeCache is complementary to other temperature-based and
structural DRAM latency reduction techniques\chI{~\cite{kim12, lee, rldram, son,
shin, lee2017design, chang2016understanding, chang2016}}. 
ChargeCache can easily be used in conjunction with \chI{any of these
techniques}.

Fourth, ChargeCache is a low-cost mechanism, which does not require any changes
to the existing DRAM chips\chs{,} and requires only small changes to the memory
controller. \chs{The low cost} makes the adoption of ChargeCache \chs{more
feasible} in future systems \chs{than other proposed mechanisms}, \chs{as these
systems} will be bottlenecked by power
consumption\chs{, and thus by complexity}~\cite{mutluresearch}. 

Overall, we believe that ChargeCache will help to significantly reduce the memory access
latency in future systems.}  \chI{To this end, to aid future research, we have 
released the source code of our ChargeCache simulator~\cite{ramulatorSharp_web, safari-github}
as part of our Ramulator releases~\cite{ramulator_web, ramulatorSharp_web}.}

 \section{Conclusion}
\label{section:conclusion}

We introduce ChargeCache, a new, low-overhead mechanism that
dynamically reduces the DRAM timing parameters for
recently-accessed DRAM rows. ChargeCache exploits two key
observations that we demonstrate in this work: 1) a
recently-accessed DRAM row has cells with high \ch{amounts} of charge
and thus \ch{it} can be accessed faster, \chs{and} 2) many applications repeatedly access
rows that are recently-accessed\ch{\chs{,} due to bank conflicts}.

Our extensive evaluations of ChargeCache on both single-core and
multi-core systems show that it provides significant performance
benefit and DRAM energy reduction at very
modest hardware overhead. ChargeCache requires no modifications
to the existing DRAM chips and occupies only a small area on the
memory controller.




We conclude that ChargeCache is a simple yet efficient mechanism to dynamically
reduce DRAM latency, which significantly improves both the performance and energy
efficiency of modern systems. \ch{We hope that our observation of the
phenomenon of row-level temporal locality and its simple exploitation to
reduce DRAM latency inspires other works to develop other new techniques to
improve memory subsystem characteristics like performance, efficiency, and
reliability.}


\section*{Acknowledgments}
\label{section:acknowledgements}

\ch{
We thank Saugata Ghose for his dedicated effort in the preparation
of this article.
We thank the reviewers and the SAFARI group members for their
feedback. We acknowledge the generous support of Google, Intel,
NVIDIA, Samsung, and VMware. This work is supported in part by NSF
grants 1212962, 1320531, and 1409723, the Intel Science and
Technology Center for Cloud Computing, and the Semiconductor
Research Corporation.}

{
\bibliographystyle{IEEEtranS}
\bibliography{refs}
}

\end{document}